\documentclass[aip,jcp,amsmath,amssymb,reprint,superscriptaddress]{revtex4-1}
\usepackage{graphicx,color}
\usepackage{epsf}
\usepackage{bm}
\usepackage{lipsum}
\def\gs{\gtrsim}

\usepackage{color}

\begin{document}

\title{
Diffusion dynamics of supercooled water modeled with the cage-jump motion and hydrogen-bond rearrangement
}

\author{Takuma Kikutsuji}
\affiliation{
Division of Chemical Engineering,
Graduate School of Engineering Science, Osaka University, Toyonaka, Osaka 560-8531, Japan
}

\author{Kang Kim}
\email{kk@cheng.es.osaka-u.ac.jp}
\affiliation{
Division of Chemical Engineering,
Graduate School of Engineering Science, Osaka University, Toyonaka, Osaka 560-8531, Japan
}
\affiliation{
Institute for Molecular Science, Okazaki, Aichi 444-8585, Japan
}

\author{Nobuyuki Matubayasi}
\email{nobuyuki@cheng.es.osaka-u.ac.jp}
\affiliation{
Division of Chemical Engineering,
Graduate School of Engineering Science, Osaka University, Toyonaka, Osaka 560-8531, Japan
}
\affiliation{
Elements Strategy Initiative for Catalysts and Batteries, Kyoto
University, Katsura, Kyoto 615-8520, Japan
}

\date{\today}

\begin{abstract}
The slow dynamics of glass-forming liquids is generally ascribed to the cage-jump motion.
In the cage-jump picture,
a molecule remains in a cage formed
by neighboring molecules, and 
after a sufficiently long time, it jumps to escape from the original position
 by cage-breaking.
The clarification of the cage-jump motion is therefore linked to
 unraveling the fundamental element of the slow dynamics.
Here, we develop 
a cage-jump model for the dynamics of supercooled water.
The caged and
 jumping states of a water molecule are introduced with respect to the
 hydrogen-bond (H-bond) rearrangement process, and describe the motion in
 supercooled states.
It is then demonstrated from the molecular dynamics
 simulation of the TIP4P/2005 model that
the characteristic length and time scales of cage-jump motions provide a
 good description of
the self-diffusion constant that is determined in turn from the long-time
 behavior of the mean square displacement.
Our cage-jump model thus enables to connect between H-bond dynamics and
 molecular diffusivity.
\end{abstract}

\maketitle

\section{Introduction}

The origin of slow dynamics observed in many supercooled liquids below
their melting temperatures is frequently 
explained utilizing the cage-effect picture.~\cite{Angell:1995dp, Gotze:1999bz}
This picture advocates that a molecule in supercooled liquids is
trapped in the cage transiently formed by neighboring molecules 
and exhibits escape jump motions due to the cage-breaking after a sufficient long time.
The cage-jump scenario also suggests intermittent molecular motions,
which can be modeled by the
continuous-time random walk using a random waiting time between cage-jumps.~\cite{Montroll:1965bv}
The cage-jump motion in glassy dynamics has been extensively addressed 
with molecular dynamics (MD) simulations~\cite{Hurley:1995gw, Doliwa:1998bg,
Yamamoto:1998jg, VollmayrLee:2004ck, Berthier:2005jh, Chaudhuri:2007bg, Saltzman:2008dp,
Candelier:2010ku,
Shiba:2012hm, Kawasaki:2013bg,
Starr:2013bs, Puosi:2013gr, Helfferich:2014ja, Hung:2019jc}
and experiments using colloidal glasses.~\cite{Pusey:1987hz,
Kasper:1998ga, Marcus:1999er, Kegel:2000dw, Weeks:2000dw,
Weeks:2002ch, Weeks:2002ji, vanMegen:2017cp}

As the temperature is decreased, the mean square displacement (MSD) exhibits
a plateau in the intermediate time scales between ballistic and
diffusive regimes, reflecting the localized motion inside the cage.
This MSD plateau value is associated with the the so-called Debye--Waller
factor 
to characterize the degree of localization.
However, it is often delicate to quantify the length and time scales of the
cage effect from an MD trajectory, which 
is continuous in space and time and is generated through thermal fluctuations.
The cage-jump model adopts a discretized view and introduces the
caged and jumping states along the dynamics of a single molecule.

Pastore \textit{et al.} have recently developed a cage-jump 
model to predict the long-time diffusivity from the short-time cage
dynamics in supercooled
liquids.~\cite{Pastore:2014hf, Pastore:2015fs, Pastore:2015da,
PicaCiamarra:2015dz, Pastore:2016iea, Pastore:2017hn}
In the study,
a trajectory of a single particle is segmented into caged 
and jumping states.
The segmentation criterion was given by the MSD plateau value.
Remarkably, the evaluations of jumping length and duration time enabled to estimate
the self-diffusion constant that is determined from the MSD long-time
behavior at any temperature.
This cage-jump modeling demonstrates that the underlying mechanism of
the molecular diffusivity is essentially governed by
the accumulation of successive cage-jump events.

The aim of this study is to develop 
a cage-jump model for supercooled water in strong connection to the
dynamics of hydrogen-bond (H-bond) network.
At normal liquid states, it has been widely accepted that a defect of 3-
or 5-coordinated H-bond plays a crucial role for characterizing the
H-bond breakage.~\cite{Laage:2006jw, Laage:2008he, Stirnemann:2012co}
By contrast, the number of defects decreases when liquid water is supercooled.
Correspondingly, the tetrahedrality of H-bond network becomes
significant, where the molecular motion is expected to be described by 
the cage-jump scenario.
Indeed, there have been various MD results showing the plateau in MSD of
supercooled water.~\cite{Sciortino:1996hk, Gallo:1996hf,
Giovambattista:2004ft, Giovambattista:2005ib, Gallo:2012cz,
Kawasaki:2017gw}
The intermittent jump motions have also been illustrated in supercooled
water by analyzing the trajectory of a single molecule.~\cite{Giovambattista:2004ft}
In particular, the connection of H-bond rearrangements
with the jump motions has been examined.

We have recently revealed that 
the H-bond lifetime $\tau_\mathrm{HB}$ depends on the temperature in
inverse proportion to the self-diffusion constant
$D$.~\cite{Kawasaki:2017gw}
This result was explained by the correlation between
H-bond breakages and translational molecular jumps. 
Moreover, we have also 
examined the pathways of hydrogen-bond breakages on the profile of the
two-dimensional potential of mean force.~\cite{Kikutsuji:2018ec}
It has been clarified that H-bonds break due to translational, rather than
rotational motions of the molecules, particularly at supercooled states.
Although these studies suggest the strong relationship between the H-bond dynamics and
molecular diffusivity in liquid water, 
the connection between the microscopic change of the H-bond
network and molecular displacement remains elusive.

The cage-jump model for supercooled water in the present work is
established by analyzing the H-bond dynamics.
In particular, 
the caged and jumping states are introduced
from the rearrangement process of four-coordinated H-bonds.
Thus, our cage-jump model does not rely on a dynamical criterion such as the MSD plateau value.
We examine how the H-bond rearrangement process links to the long-time
diffusivity.

\section{Model and simulations}
\label{sec:model}

MD simulations were performed using the TIP4P/2005 water model.~\cite{Abascal:2005ka}
All the simulations in this work were performed with the GROMACS2016.4 package.~\cite{Hess:2008db, Abraham:2015gj}
The phase diagram of the TIP4P/2005 suerpcooled water was determined in
Refs.~\onlinecite{Singh:2016bu, Biddle:2017bb}.
The temperature crossing the Widom line in the $\rho$-$T$ phase diagram
is $T_\mathrm{L}\approx 210$ K at 1 $\mathrm{g/rm^3}$.
Furthermore, the mode-coupling glass transition is estimated as
$T_\mathrm{C}\approx 190$ K  at 1 $\mathrm{g/rm^3}$ in
Refs.~\onlinecite{DeMarzio:2016hl, DeMarzio:2017kh}.
Recent MD simulations have reported that the divergence of
the structural relaxation time occurs at $T_\mathrm{g} \approx 136$ K under the
constant 1 bar condition.~\cite{Saito:2018dn, Saito:2019ht}
As described next, we examined the systems close to $T_\mathrm{L}$, while our
simulation temperatures are above $T_\mathrm{C}$ and $T_\mathrm{g}$.

The mass density was fixed at 1 $\mathrm{g/cm^3}$ and the simulation
system contained $N=8,000$ molecules in a cubic box with the periodic
boundary conditions.
The cell length was approximately $L=6.2$ nm.
The investigated temperatures were $T =$ 350, 320,
300, 280, 260, 240, 220, 210, 200,
and 190 K. 
At each temperature, the system was equilibrated for 10 ns in the $NVT$
ensemble, followed by a production run in $NVE$ for 20 ns.
Other trajectories of 100 ns were generated for 
MSD and H-bond correlation function (see
the definitions below) at temperatures 200 K and 190 K.
A time step of 1 fs was used. 
As demonstrated below, 
this trajectory duration is larger than the H-bond lifetime
$\tau_\mathrm{HB}$ at all the temperatures examined.
Furthermore, aging effects have not been detected in the
course of MD simulations.
The atomic coordinates were stored at 0.2 ps intervals, which were used 
for the analyses presented below.
This interval was chosen as a 
time scale slightly larger
than that of
libration motions ($\sim 0.1$ ps).

The MSD, $\langle \delta r^2(t)\rangle=\langle\sum_{i=1}^N|\Delta \bm{r}_i(t)|^2
\rangle / N$, was calculated to quantify the self-diffusion constant $D$.
Here, $\Delta \bm{r}_i(t)$ represents the displacement vector of an O
atom of the molecule $i$ between two times 0 and $t$.
The results at various temperatures are 
shown in the inset of Fig.~\ref{fig:jmsd}(b).
At temperatures below 280 K, 
a plateau becomes noticeable during the intermediate times between
ballistic and diffusive regimes, indicating 
the cage effect.
The self-diffusion constant $D$ was determined from 
the long-time behavior of the MSD, $D=\lim_{t\to\infty}\langle \delta r^2(t)\rangle/6t$.
The ratio of the Lennard--Jones diameter of TIP4P/2005 model and the unit cell is 
$\sigma/L\approx 0.05$, which is sufficiently small to eliminate the
finite-size effect on the diffusion constant.~\cite{Yeh:2004gs}

The H-bond was defined using geometric variables between two water molecules.
We adopted O-O distance $R$ and OH-O angle $\beta$.~\cite{Kumar:2007bs}
Two water molecules are considered H-bonded if the distance-angle relationship
meets the condition, ($R$, $\beta$) $\leq$ (0.34 nm, 30$^\circ$).
The H-bond correlation function 
$
c(t) = \langle h(0) h(t)\rangle/ \langle h(0) \rangle
$
was calculated 
with the H-bond indicator $h(t)$ at a time $t$.~\cite{Luzar:1996gw, Luzar:1996gx}
It analyzes `history-independent' H-bond correlations,
in the sense that $h(t)$ is evaluated only from the configuration at
time $t$ without taking into account the reformation of the H-bond in the
interval between times $0$ and $t$.
The H-bond lifetime $\tau_\mathrm{HB}$ was then determined from $c(t)$ by fitting it to
the stretched-exponential function $\exp[-(t/\tau_\mathrm{HB})^\beta]$.

\begin{figure}[t]
\centering
\includegraphics[width=0.5\textwidth]{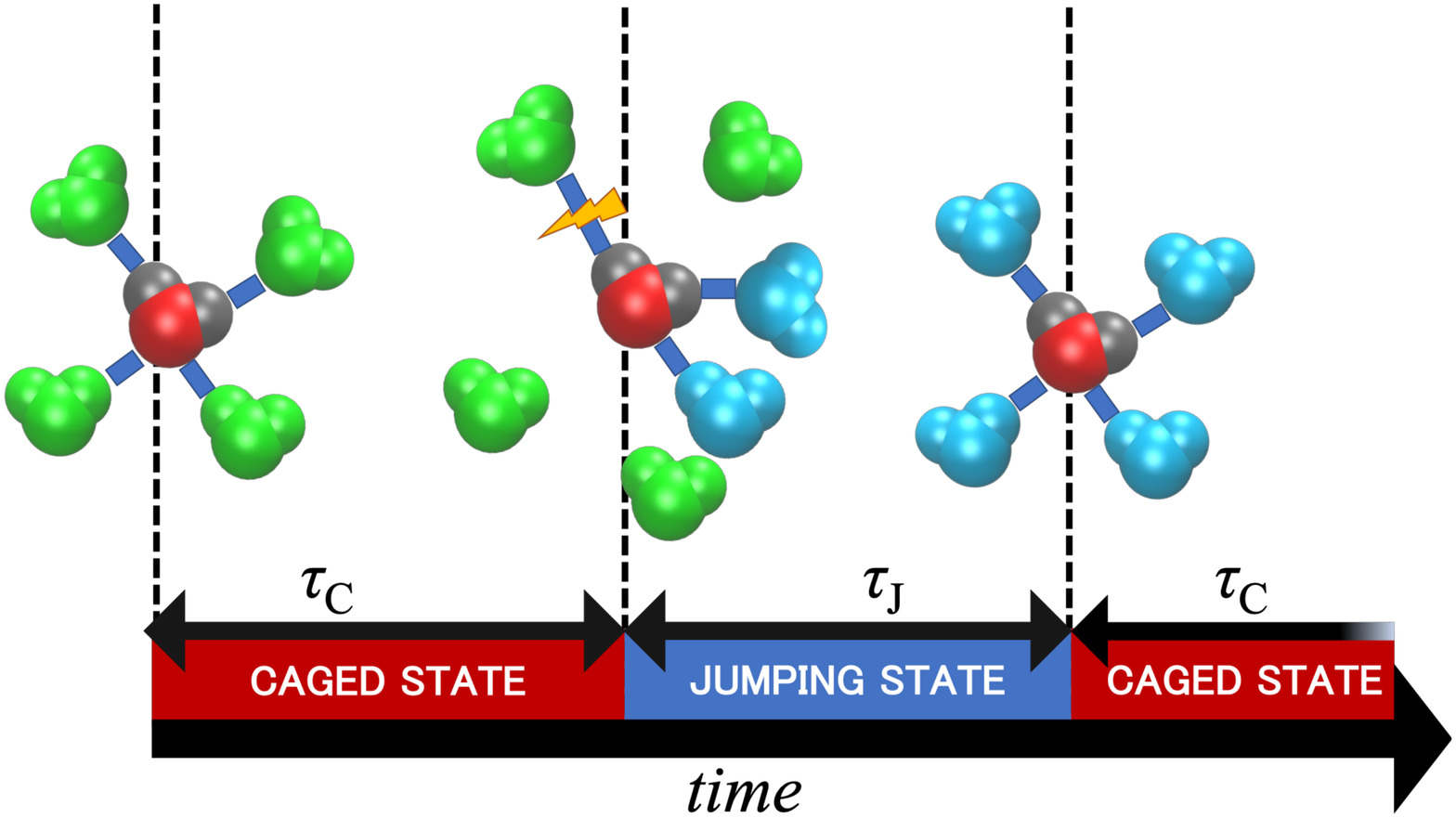}
\caption{Schematic illustration to distinguish the caged (C) and jumping (J) 
states for a tagged water molecule (gray color).
The tagged water molecule are H-bonded with neighboring four water
 molecules (green color), during which the state is labeled as a C
 state.
At a certain point of time, all of four H-bonds are broken and the C state is switched
 to the J state.
After a period of time, the tagged molecule are H-bonded with completely
different four water molecules (cyan color), at which the state is switched to the
 next C state.
The duration times of the J and C states are denoted as
 $\tau_\mathrm{J}$ and $\tau_\mathrm{C}$, respectively.}
\label{fig:switching}
\end{figure}

\begin{figure}[t]
\centering
\includegraphics[width=0.45\textwidth]{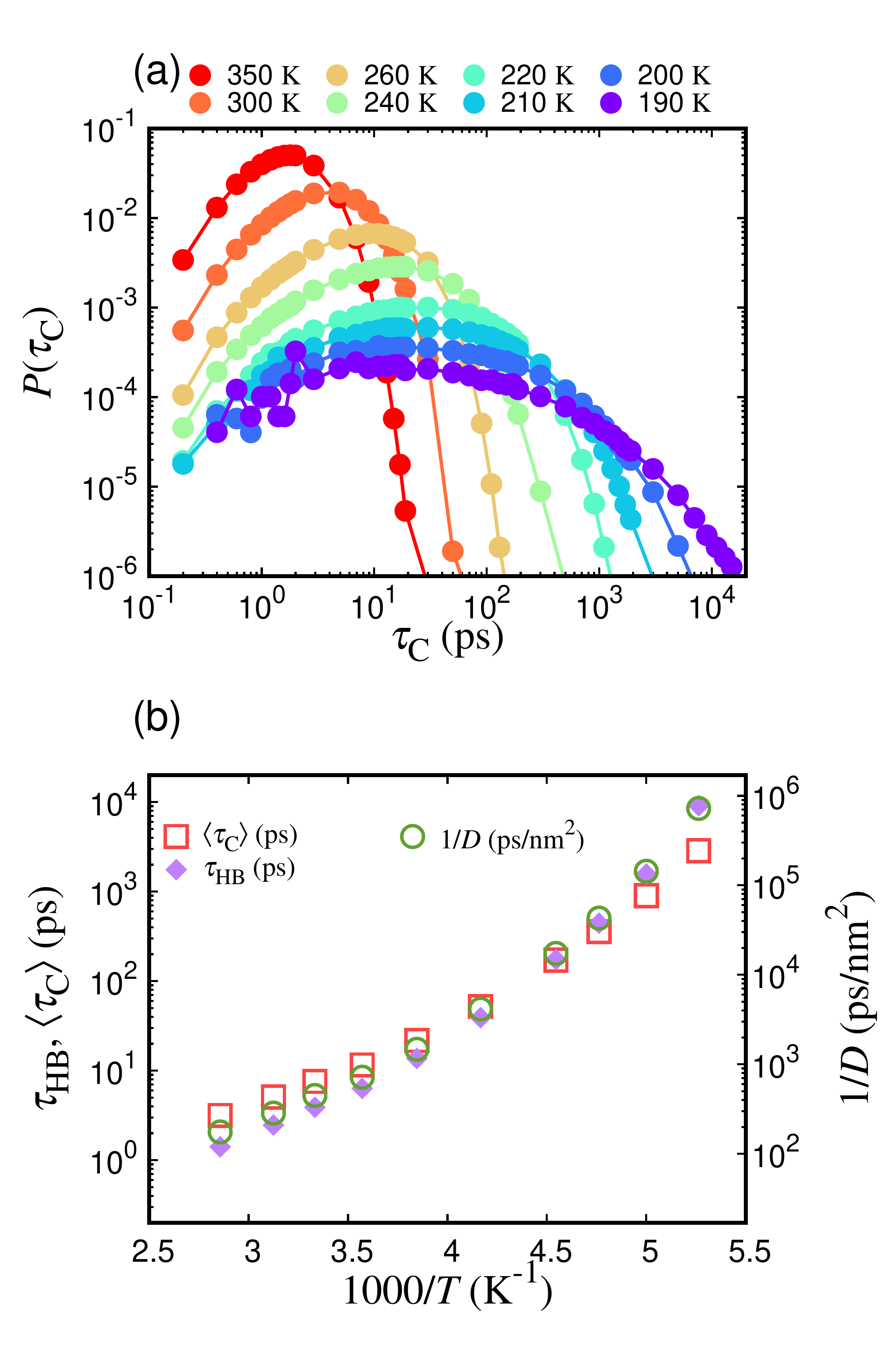}
\caption{
(a) Distribution of the 
duration time of the C state,
 $P(\tau_\mathrm{C})$, at the temperatures examined.
(b) Average duration time $\langle
 \tau_\mathrm{C}\rangle$ (left axis), H-bond lifetime $\tau_\mathrm{HB}$
 (left axis), and inverse
 of self-diffusion constant $D^{-1}$ (right axis), as a function of the inverse of
 the temperature, $1000/T$.
}
\label{fig:tau_C} 
\end{figure}

\begin{figure}[t]
\centering
\includegraphics[width=0.45\textwidth]{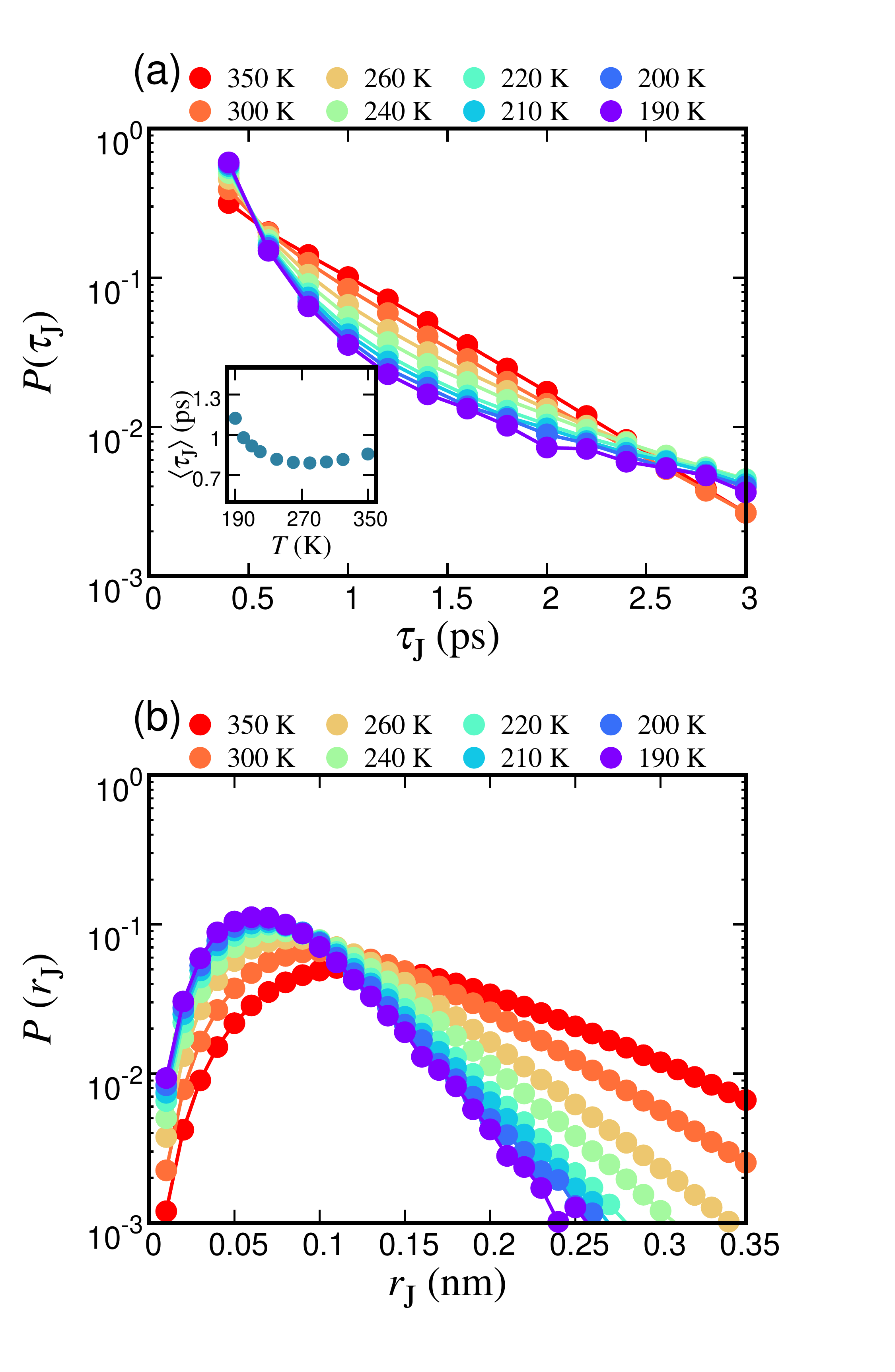}
\caption{
(a) Distribution of the 
duration time of the J state,
 $P(\tau_\mathrm{C})$, at the temperatures examined.
Inset: temperature dependence of average jump time $\langle \tau_\mathrm{J}\rangle$.
(b) Distribution of the jumping length during the jumping state,
 $P(r_\mathrm{C})$, at the temperatures examined.
}
\label{fig:tau_J} 
\end{figure}

We classify the time course of each water molecule into two states. 
One
is called caged (C) state, where the tagged water molecule is initially H-bonded
to other four water molecules. 
The schematic illustration of the J and C states
is given in Fig.~\ref{fig:switching}.
Since an H-bond is of finite lifetime,
the four H-bonds with the tagged molecule are all broken at a certain
time. 
This time is set to the start of the jumping (J) state. 
The next C
state then begins at the formation of four H-bonds with water molecules
that are totally different from those in the previous C state.
The complete changes of the H-bond partners is the condition of
transition from one C
state to the next.
The end time of a C state is when the four H-bonds
are first broken, and the start time is when new, four bonds are
formed. 
The adjacent C states are bridged by a J state, and by
definition, a C state may be of 1, 2, 3 or 4 H-bonds and a J state may
experience the reformation of an H-bond that was present in the previous
C state. 
When an H-bond in the previous C state reforms after all the
four bonds are once broken, the tagged molecule is still in the J state.
The duration times of the C and J states are denoted as $\tau_\mathrm{C}$ and $\tau_\mathrm{J}$, respectively.
We also quantified the displacement vectors of the O atom of the
molecule $i$ during the J state, which is represented as $\Delta \bm{r}^{\mathrm{J}}_i(\theta)$.
Here, $\theta$ is the counter for molecule $i$ to stay 
at J states from
the initial time of the trajectory.
Furthermore, many $\tau_\mathrm{C}$ and $\tau_\mathrm{J}$ were obtained for the single-molecule
trajectory of each water molecule. 
The averages of $\tau_\mathrm{C}$ and $\tau_\mathrm{J}$ over
all the single-molecule trajectories are denoted as $\langle \tau_\mathrm{C}\rangle$
and $\langle \tau_\mathrm{J}\rangle$, respectively. 
On the other hand, the sum of $\tau_\mathrm{J}$ along a
single trajectory was obtained and its ratio to the total length of that
trajectory was also determined. 
The average of this ratio over all the
single-molecule trajectories is then called $\rho_\mathrm{J}$.
Due to the difference in the order of averaging, $\rho_\mathrm{J}$ is in principle different
from 
$\langle \tau_\mathrm{J}\rangle/(\langle \tau_\mathrm{C}\rangle +
\langle \tau_\mathrm{J}\rangle)$; this point will be examined at the end
of Sec.~\ref{sec:results}.
These quantities provide time coarse-grained information 
filtering out the thermal fluctuations within the J states as well as the libration motions.

\section{Results and discussion}
\label{sec:results}

\begin{figure}[t]
\centering
\includegraphics[width=0.4\textwidth]{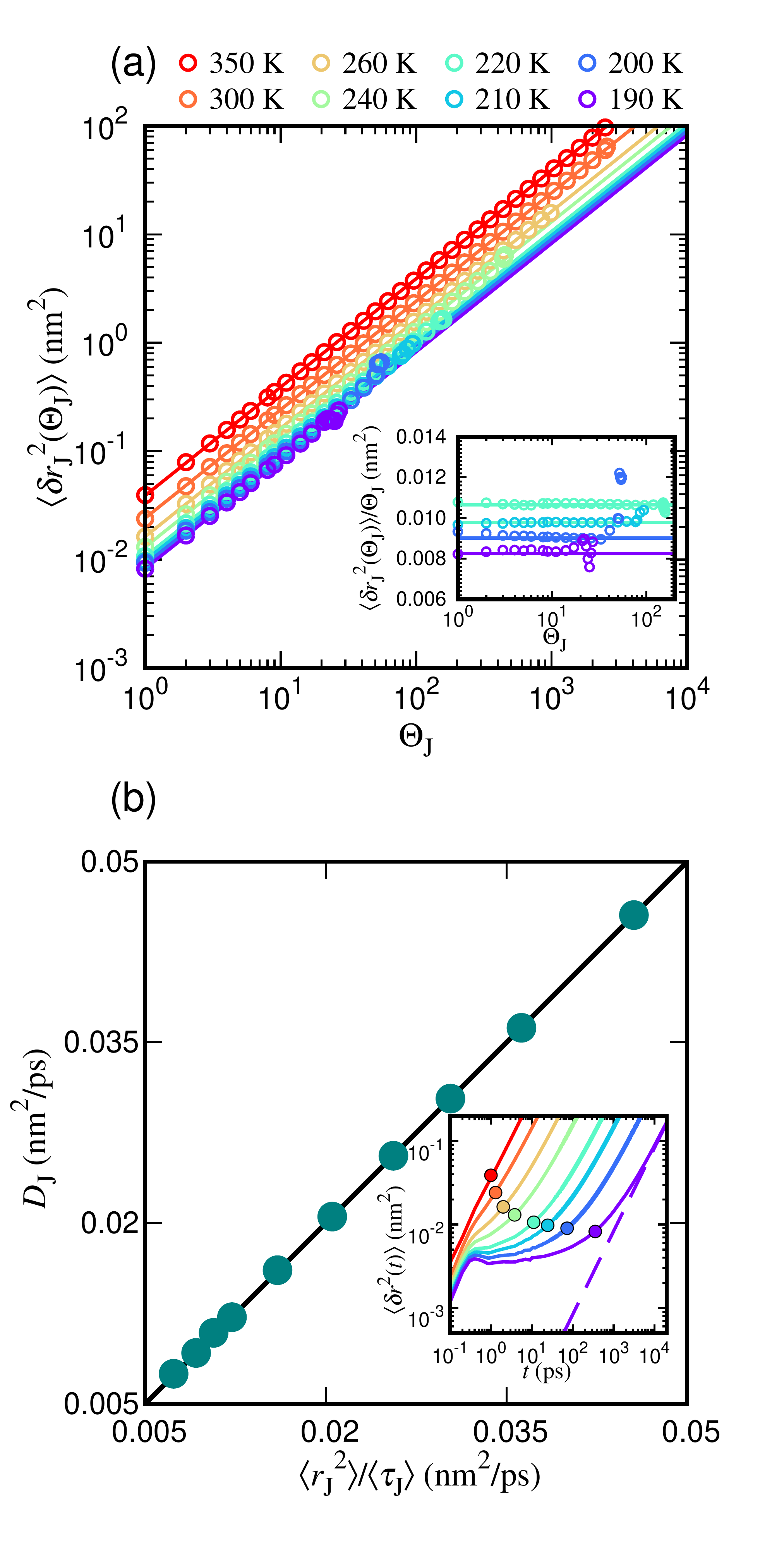}
\caption{
(a) Jumping mean square displacement (JMSD), $\langle \delta
 r^2_\mathrm{J}(t)\rangle$, as a function of the number
 of jumps $\Theta_\mathrm{J}$ at the temperatures examined.
Inset: $\langle \delta r^2_\mathrm{J}(t)\rangle/
 \Theta_\mathrm{J}$ as a function of $\Theta_\mathrm{J}$ at temperatures
below $T=200$ K.
The straight lines indicate the diffusion behavior, of which
 slope gives a jumping self-diffusion constant $D_\mathrm{J}$.
(b) Jumping self-diffusion constant $D_\mathrm{J}$ vs. $\langle
 r_\mathrm{J}^2\rangle/ \langle \tau_\mathrm{J}\rangle$ obtained from
 $P(\tau_\mathrm{J})$ and $P(r_\mathrm{J})$.
The black line represents $D_\mathrm{J} =\langle r_\mathrm{J}^2\rangle/ \langle \tau_\mathrm{J}\rangle$.
Inset: Mean square displacement (MSD), $\langle \delta r^2(t)\rangle$, at the temperatures examined.
Dashed line represents the long-time asymptote $6Dt$ at $T=190$ K.
Points indicate values of the average jump length $\langle
 r_\mathrm{J}^2\rangle$ at each temperature.
}
\label{fig:jmsd} 
\end{figure}

\begin{figure}[t]
\centering
\includegraphics[width=0.4\textwidth]{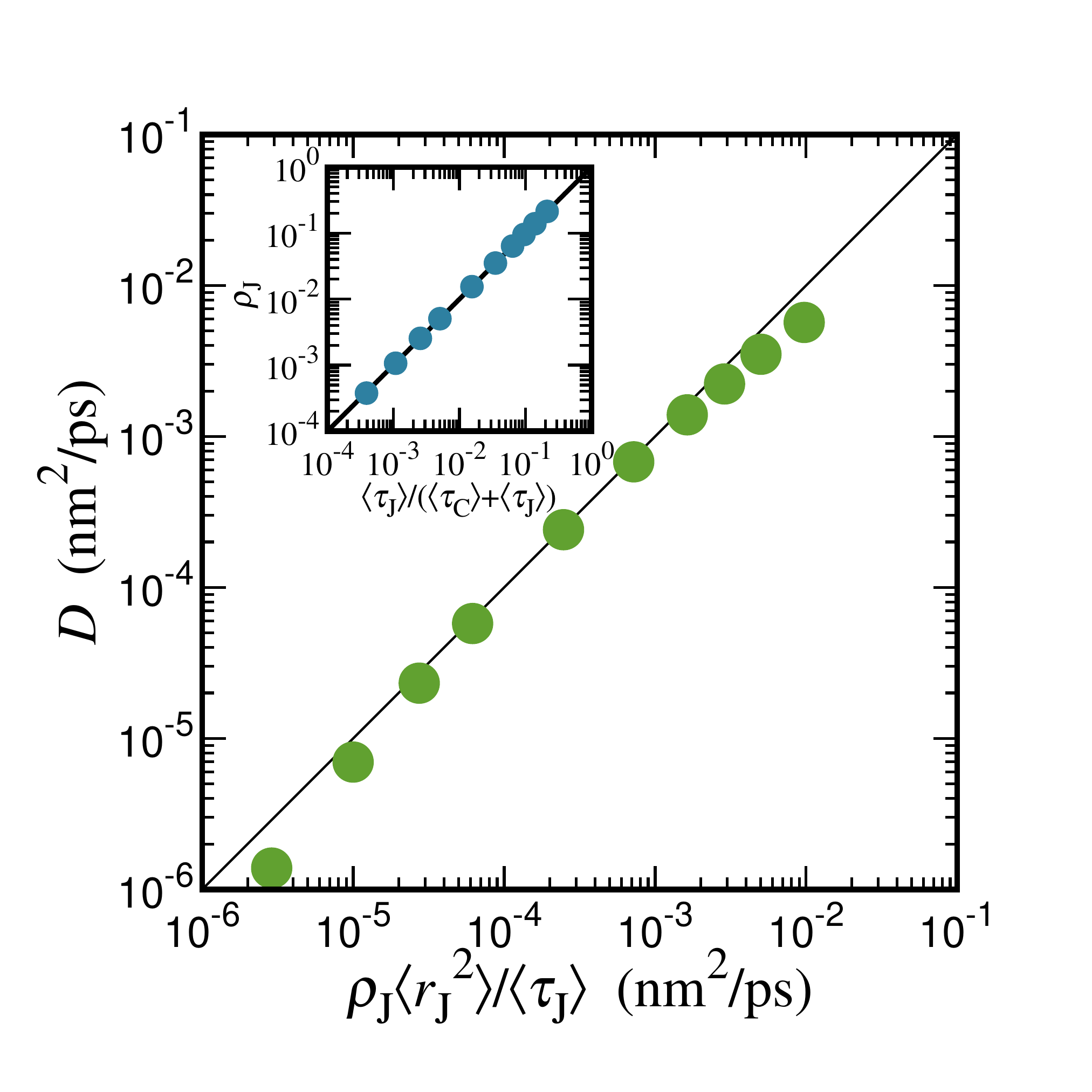}
\caption{
Self-diffusion constant $D$ vs. 
the estimate from the cage-jump model $\rho_\mathrm{J}\langle
 r_\mathrm{J}^2\rangle / \langle \tau_\mathrm{J}\rangle$.
The black line represents $D=\rho_\mathrm{J}\langle
 r_\mathrm{J}^2\rangle / \langle \tau_\mathrm{J}\rangle$.
Inset: Ratio of jumping state $\rho_\mathrm{J}$ vs. the ratio of average
jumping time $\langle
 \tau_\mathrm{J}\rangle/ (\langle \tau_\mathrm{C}\rangle +
 \langle\tau_\mathrm{J}\rangle)$.
The black line represents $\rho_\mathrm{J}=\langle
 \tau_\mathrm{J}\rangle/ (\langle \tau_\mathrm{C}\rangle +
 \langle\tau_\mathrm{J}\rangle)$.
} 
\label{fig:diffusion}
\end{figure}

Figure~\ref{fig:tau_C}(a)
shows the distribution of the duration time of the C state,
$P(\tau_\mathrm{C})$, at the temperatures examined.
The peak of $P(\tau_\mathrm{C})$ appears at around 10 ps at 300 K,
which shifts to time scale of 50 ps at 190 K.
In addition, the distribution is gradually extended to slower time
scales with decreasing the temperature.
The temperature dependence of the average duration time $\langle
\tau_\mathrm{C}\rangle$ is plotted in Fig.~\ref{fig:tau_C}(b).
In comparison, the temperature dependence of $\tau_\mathrm{HB}$ and
$D^{-1}$ is also plotted in Fig.~\ref{fig:tau_C}(b).
It is demonstrated that the time scales of $\langle
\tau_\mathrm{C}\rangle$ is akin to $\tau_\mathrm{HB}$, although the
temperature dependence is slightly different.
Note that the mean value of H-bond number depends on the temperature 
ranging from 3.62 at 300 K to 3.97 at 190 K, presumably resulting in the
difference between $\langle\tau_\mathrm{C}\rangle$ and $\tau_\mathrm{HB}$.
Furthermore, 
the intimate connection between self-diffusion constant $D$ and H-bond
lifetime $\tau_\mathrm{HB}$ is clarified in Fig.~\ref{fig:tau_C}(b),
which is equivalent to the previous demonstration,
$D\propto{\tau_\mathrm{HB}}^{-1}$, in TIP4P/2005 supercooled water.~\cite{Kawasaki:2017gw}
The roles of $\langle\tau_\mathrm{C}\rangle$ and
$\tau_\mathrm{HB}$ for the cage-jump model will be discussed later.

Figure~\ref{fig:tau_J}(a) shows the distribution of the duration time
of the J state, $P(\tau_\mathrm{J})$.
Contrary to $P(\tau_\mathrm{C})$ in Fig.~\ref{fig:tau_C},
the temperature dependence of $P(\tau_\mathrm{J})$ is much smaller.
This causes the very weak temperature dependence of the average jumping time
$\langle \tau_\mathrm{J}\rangle\approx  1$ ps, as seen in the inset of Fig.~\ref{fig:tau_J}(a).
Figure~\ref{fig:tau_J}(b) shows the distribution of the jumping length
during the J state, $P(r_\mathrm{J})$ with $r_\mathrm{J}=|\Delta \bm{r}^{\mathrm{J}}_i|$.
This demonstrates that 
the peak position length scale of $P(r_\mathrm{J})$ becomes smaller as the temperature is decreased.
Furthermore, the exponential decay of $P(r_\mathrm{J})$ beyond the peak
position is clearly observed at each temperature.
Our results of $P(\tau_\mathrm{J})$ and $P(r_\mathrm{J})$ are compatible
with those of supercooled liquid using simple potentials.~\cite{Pastore:2014hf}
Note that these distributions $P(\tau_\mathrm{J})$ and
$P(r_\mathrm{J})$ are sufficiently converged when 
plenty of jump events are accumulated, requiring
times exceeding $\langle \tau_\mathrm{C}\rangle$ and $\tau_\mathrm{HB}$.

We calculated the jumping mean square displacement (JMSD),
\begin{equation}
\langle \delta r_\mathrm{J}^2(\Theta_\mathrm{J})\rangle = \frac{1}{N_\mathrm{J}}
\sum_{i=1}^{N_\mathrm{J}}\sum_{\theta=1}^{\Theta_\mathrm{J}} |\Delta
\bm{r}^\mathrm{J}_i(\theta)|^2, 
\end{equation}
where $\Theta_\mathrm{J}$ is the total number of J states during an MD
trajectory of a single water molecule.
In addition, $N_\mathrm{J}$ denotes the number of molecules
exhibiting $\Theta_\mathrm{J}$ jumps in the course of the trajectory.
Figure~\ref{fig:jmsd}(a) shows that the JMSD $\langle \delta
r_\mathrm{J}^2(\Theta_\mathrm{J})\rangle$ 
is linear with $\Theta_\mathrm{J}$ at each temperature.
However, we observed some deviations in the JMSD, particularly for
larger $\Theta_\mathrm{J}$ at lower temperatures (see the inset of Fig.~\ref{fig:jmsd}(a)).
These deviations from the linearity at large $\Theta_\mathrm{J}$ are due
to the fact that $N_\mathrm{J}$ becomes much
smaller than $N$, leading to the insufficient ensemble average over the
molecules.
At 190 K, for example, $N_\mathrm{J}$ becomes smaller than $N$ at 
$\Theta_\mathrm{J} \gs 10$.
The average self-diffusion constant $D_\mathrm{J}$ of successive jumping
events is determined from the 
relation, $D_\mathrm{J}=\lim_{\Theta_\mathrm{J}\to\infty} \langle \delta
r_\mathrm{J}^2(\Theta_\mathrm{J})\rangle / (\Theta_\mathrm{J}\langle
\tau_\mathrm{J}\rangle)$, where the linear fit was done by excluding the $N_\mathrm{J}< N$ regions.
Figure~\ref{fig:jmsd}(b) demonstrates that the second-order moment
$\langle r_\mathrm{J}^2\rangle$ of the distribution $P(r_\mathrm{J})$ 
provides the good description of $D_\mathrm{J}$ at each temperature.
As discussed in Ref.~\onlinecite{Pastore:2014hf}, 
these features of the JMSD and $D_\mathrm{J}$ indicate that the
diffusion process of a single molecule can be described by the
random walk with independent jumps, which is
characterized by $\langle r_\mathrm{J}^2\rangle$.
Furthermore, the decrease in $D_\mathrm{J}$ with the temperature reduction is attributed to the
corresponding decrease in the jumping length scale $\langle r_\mathrm{J}^2\rangle$.
The inset of Fig.~\ref{fig:jmsd}(b) illustrates the location of $\langle
r_\mathrm{J}^2\rangle$ in the MSD.
At each temperature, the value of $\langle r_\mathrm{J}^2\rangle$
slightly exceeds beyond the MSD plateau.
This observation indicates the validity of our modeling for cage-jump motions.

The relevance of the cage-jump modeling 
is examined in Fig.~\ref{fig:diffusion} by plotting the relationship
between the self-diffusion constant $D$ and $\rho_\mathrm{J}D_\mathrm{J}=\rho_\mathrm{J}\langle
r_\mathrm{J}^2\rangle/\langle \tau_\mathrm{J}\rangle$.
Note that the inset of Fig.~\ref{fig:diffusion} demonstrates 
the ratio of the J state $\rho_\mathrm{J}$ 
is essentially equal to
the ratio of the average
jumping time
$\langle \tau_\mathrm{J}\rangle/ (\langle \tau_\mathrm{C}\rangle + \langle
\tau_\mathrm{J}\rangle)$.
A similar result was also reported in Ref.~\onlinecite{Pastore:2014hf}.
It should be noted that $\rho_\mathrm{J}$ is obtained by first analyzing each
single-molecule trajectory and then taking an average over all the
single-molecule trajectories, while $\langle \tau_\mathrm{J} \rangle$ and
$\langle \tau_\mathrm{C}\rangle$ are computed
without distinguishing the trajectories of distinct water molecules. 
$\rho_\mathrm{J} = \langle \tau_\mathrm{J}\rangle/( \langle
\tau_\mathrm{C}\rangle + \langle \tau_\mathrm{J}\rangle)$ thus implies the
validity of a mean-field-type view in that the average over the
single-molecule trajectories can be determined without taking into
account the differences among the trajectories.
Figure~\ref{fig:diffusion} shows that
$\rho_\mathrm{J}D_\mathrm{J}$ 
from our cage-jump model is a good indicator 
for the self-diffusion constant $D$, although slight deviations are
apparent at 190 K and higher temperatures above 300 K.
As shown in Fig.~\ref{fig:tau_C}(b), the inverse of the self-diffusion constant $D^{-1}$ is
strongly coupled with $\tau_\mathrm{HB}$, which becomes slightly larger
than $\langle \tau_\mathrm{C}\rangle$.
When $\langle \tau_\mathrm{J}\rangle\ll \langle\tau_\mathrm{C}\rangle$, 
$\rho_\mathrm{J}D_\mathrm{J}$ is approximated by $\langle
r_\mathrm{J}^2\rangle / \langle \tau_\mathrm{C} \rangle$, 
which holds particularly
at lower temperatures.
Thus, the difference between $\langle \tau_\mathrm{C}\rangle$ and
$\tau_\mathrm{HB}$ results in the small difference between $D$ and $\rho_\mathrm{J}D_\mathrm{J}$ at 190 K.
The deviations at high temperatures are attributed to the observation that
the MSD plateau was not well developed, as shown in the inset of
Fig.~\ref{fig:jmsd}(b).
In fact, at higher temperatures, the cage
structures are weakened and water molecules immediately diffuse
without exhibiting intermittent cage-jump motions.

\section{Conclusions}
\label{sec:conclusions}

In this paper, we have developed 
a cage-jump model for the diffusion in supercooled water. 
Unlike the scheme proposed by Pastore
\textit{et al.},~\cite{Pastore:2014hf, Pastore:2015fs, Pastore:2015da,
PicaCiamarra:2015dz, Pastore:2016iea, Pastore:2017hn}
we classify the trajectory of a single water molecule into the caged and jumping states from
the analysis of H-bond rearrangements.
The quantification of the average length and time scales of the jumping
state enabled to predict the self-diffusion constant $D$ that is
determined in principle from the long-time MSD behavior. 
We have thus succeeded in connecting the H-bond
dynamics and the molecular diffusivity 
through the cage-jump events.
This cage-jump event can be regarded as an element of the
collective motions, which are often visualized by string-like
motions.~\cite{Giovambattista:2004ft}
In fact, the time scale is $\langle \tau_\mathrm{J}\rangle\approx$ 1 ps,
whereas that of collective motions is typically characterized by the
time of the last stage in the MSD plateau.

Our cage-jump model gives an estimate of $D$ 
when the caged and jumping states are identified from an MD trajectory,
without extensive MSD evaluation to the diffusive regime.
In fact, the diffusion asymptote $6Dt$ is observed at times much larger
than $\tau_\mathrm{HB}$, as shown in the inset of Fig.~\ref{fig:jmsd}(b).
However, particularly at lower temperatures, the duration time of the C state $\langle
\tau_\mathrm{C}\rangle$ becomes slightly smaller than the H-bond
lifetime $\tau_\mathrm{HB}$.
This causes the small difference between $D$ and
$\rho_\mathrm{J}D_\mathrm{J}$, particularly at 190 K, as observed in Fig.~\ref{fig:diffusion}.
The local structure changes from high-density liquid
to low-density liquid with decreasing the temperature below the so-called Widom line
($T_\mathrm{L}\approx 210$ K).
The tetrahedral order becomes higher and the number of the
defect correspondingly decreases in the low-density liquid state, where
the hydrogen-bond break needs more activation energy, and correspondingly
the H-bond network 
rearranges over a wider range in space.
In contrast, our cage-jump model is constructed based on the information on the first
nearest neighbor shell only.
This leads to the underestimation of $\langle\tau_\mathrm{C}\rangle$
compared with $\tau_\mathrm{HB}$.
A possible refinement is thus to 
incorporate order parameters for H-bond network such as 
the local structure index~\cite{Shiratani:1996ea, Shiratani:1998fq,
Cuthbertson:2011fb, Shi:2018fl, Saito:2018dn},
which focus on the second nearest neighbor.
In this respect, further study is currently undertaken toward the 
appropriate inference of the
self-diffusion constant $D$, particularly at much deeper supercooled
states inside the so-called no man's land region.~\cite{Shi:2018gu,
Shi:2018is, Saito:2018dn, Ni:2018ie, Saito:2019ht}.
It is also worthy to apply 
our cage-jump model to various water models to give deeper insight into the
role of the H-bond breakage on the molecular diffusivity in
supercooled water.

\begin{acknowledgments}
The authors thank S. Saito and T. Kawasaki for helpful discussions.
This work was supported by JSPS KAKENHI Grant Numbers JP18H01188 (K.K.),
 JP15K13550 (N.M.), and JP19H04206 (N.M.).
This work was also supported in part by 
the Post-K Supercomputing Project and the Elements
Strategy Initiative for Catalysts and Batteries from the Ministry of
 Education, Culture, Sports, Science, and Technology.
The numerical calculations were performed at Research Center of Computational
Science, Okazaki Research Facilities, National Institutes of Natural Sciences, Japan.
\end{acknowledgments}

%aipnum4-2.bst 2019-01-14 (MD) hand-edited version of apsrev4-1.bst

\end{document}